\documentclass[aps, %
twocolumn, %
preprintnumbers, %
floatfix, %
superscriptaddress, %
groupedaddress, %
prl
]{revtex4} %

\usepackage{amssymb}%
\usepackage{amsmath}%
\usepackage{graphicx}%

\def\e{{\rm e}}

\def\d{\partial}

\newcommand{\be}{\begin{equation}}
\newcommand{\ee}{\end{equation}}
\newcommand{\bea}{\begin{eqnarray}}
\newcommand{\eea}{\end{eqnarray}}
\newcommand{\bg}{\begin{gather}}
\newcommand{\eg}{\end{gather}}
\newcommand{\bseq}{\begin{subequations}}
\newcommand{\eseq}{\end{subequations}}

\renewcommand{\Im}{\mathop{\rm Im}\nolimits}
\renewcommand{\Re}{\mathop{\rm Re}\nolimits}

\newcommand{\bra}[1]{\langle #1 |}
\newcommand{\ket}[1]{| #1 \rangle}

\newcommand{\bpm}{\begin{pmatrix}}
\newcommand{\epm}{\end{pmatrix}}

\makeatletter

\def\compoundrel#1\over#2{\mathpalette\compoundreL{{#1}\over{#2}}}
\def\compoundreL#1#2{\compoundREL#1#2}
\def\compoundREL#1#2\over#3{\mathrel
         {\vcenter{\hbox{$\m@th\buildrel{#1#2}\over{#1#3}$}}}}
\makeatother

\allowdisplaybreaks[4]

\begin{document}

\preprint{CERN-PH-TH/2007-109}

\title{Unstable Semiclassical Trajectories in Tunneling}
\author{D.G.~Levkov}
\affiliation{Institute for Nuclear Research of the Russian Academy of
  Sciences, 60th October Anniversary prospect 7a, Moscow 117312,
  Russia}

\author{A.G.~Panin}
\affiliation{Institute for Nuclear Research of the Russian Academy of
  Sciences, 60th October Anniversary prospect 7a, Moscow 117312, Russia}
\affiliation{Moscow Institute of Physics and Technology, 
Institutskii per. 9, Dolgoprudny 141700, Moscow Region, Russia}

\author{S.M.~Sibiryakov}
\affiliation{Theory Group, Physics Department, CERN, CH-1211 Geneva 23,
  Switzerland}
\affiliation{Institute for Nuclear Research of the Russian Academy of
  Sciences,  60th October Anniversary prospect 7a, Moscow 117312, Russia}

\begin{abstract}
  Some tunneling phenomena are described, in the semiclassical
approximation, by unstable complex
  trajectories. We develop a systematic
  procedure to stabilize the trajectories and to calculate the
  tunneling probability, including both the suppression exponent and
  prefactor. We find that the instability of tunneling solutions
  modifies the power--law dependence of the prefactor on $\hbar$ as
  compared to the case of stable solutions.
\end{abstract}

\maketitle

Tunneling in systems with many degrees of freedom has been a subject
of continuous theoretical research for the last decades 
\cite{bohigas}. The interest is
heated up by the recent  
experimental observations~\cite{Dembowski,Hensinger,Steck} of
non--trivial dynamical properties of multidimensional tunneling. 
In the semiclassical framework tunneling is described as
motion of the system along a complex trajectory -- solution to the
classical equations of motion analytically continued to the complex
values of coordinates and/or time
\cite{Miller,recentreviews}. Given the
complex trajectory, one calculates the tunneling probability at
small values of the Planck constant $\hbar$,
\be
\label{main}
{\cal P}=A\e^{-F/\hbar}\;,
\ee
where $F$ and $A$ are the suppression exponent and prefactor,
respectively. 

Recently a new, intrinsically multidimensional, mechanism of tunneling
has been discovered \cite{Bezrukov:2003yf,Takahashi:Ikeda}. It differs
qualitatively from the well-studied case of direct tunneling, where
complex trajectories connect the in- and out- regions of the phase
space. The new mechanism generically occurs in the situation when the
total energy of the system exceeds the height of the potential barrier
separating the in- and out- states but, still, the process remains
exponentially suppressed (dynamical tunneling). The complex
trajectories in the
new mechanism end up on a real unstable periodic orbit lying on the
boundary  between the in- and out- regions. We call this orbit
``sphaleron'';  its instability implies that the tunneling  
trajectories are also unstable. The above behavior of complex
trajectories leads to the physical picture of tunneling as a two-step
process \cite{Bezrukov:2003yf,Bezrukov:2003er}: 
formation of the sphaleron
and its decay into the out-region. The latter step is not described by
the tunneling trajectory; on the other hand, it does not involve
exponential suppression, as the decay of the sphaleron proceeds
classically. 
This tunneling mechanism is generic and has been found in several
quantum mechanical \cite{Levkov:2007ce,Takahashi:Ikeda2} and field
theoretical \cite{Bezrukov:2003er} models. It is natural to call the
new mechanism ``sphaleron--driven'' tunneling~\cite{Wilkinson}. 

Tunneling via unstable semiclassical solutions raises a
number of issues. First, search for unstable trajectories is 
problematic from the numerical point of view. Second, even if one
finds the tunneling trajectory which tends to the sphaleron as $t\to
+\infty$, a problem remains to describe semiclassically the subsequent
decay of the sphaleron orbit. Yet another issue is
the calculation of the prefactor $A$. In the case of direct
tunneling this calculation involves the analysis of 
linear perturbations around the tunneling trajectory. This procedure
is not applicable in the sphaleron--driven case, when the perturbations
destroy the tunneling solution  completely.

In this Letter we systematically develop a general method to solve 
the above problems. The
idea is to introduce a constraint into the path integral for the
tunneling amplitude. This modifies the equations of motion
in such a way that tunneling trajectories are pushed away from the
sphaleron. The modification is governed by a
regularization parameter $\epsilon$. At $\epsilon>0$ the
semiclassical solutions interpolate between the in-- 
and out--regions 
and are stable. They describe the whole
two-stage process of sphaleron-driven tunneling. Integration over the
constraint corresponds to taking the limit $\epsilon\to +0$.
The original unstable trajectory is recovered from the regularized
solutions in this limit. We find expressions for the suppression
exponent and prefactor of the sphaleron-driven tunneling
probability
in a closed form. 
Our method is based on the ideas of \cite{Bezrukov:2003yf}.

Our analysis reveals, among other things, one
universal feature: the prefactor $A$ in the probability
of tunneling via unstable
trajectories gets suppressed by an additional factor $\hbar^{1/2}$ as
compared to the case of direct tunneling. This, in particular, implies
non-trivial properties of  
the transition between the
direct and sphaleron--driven tunneling regimes.

Remarkably, the
method of $\epsilon$-regularization can be used to 
deform real solutions corresponding to purely classical motion
into trajectories describing tunneling \cite{Bezrukov:2003yf,Levkov:2007ce}.
This makes the method efficient for finding and classifying complex solutions
in the case of chaotic  tunneling \cite{Levkov:2007ce}. We stress,
however, that the phenomenon of sphaleron--driven tunneling is
unrelated to chaos and was observed both in chaotic and regular
systems.
 
While our approach is completely general, for concretness, 
we illustrate it using a 
two--dimensional model with
the Hamiltonian 
\be
\label{Ham}
H=(p_x^2+p_y^2+\omega^2 y^2)/2+\exp[-(x+y)^2/2]\;.
\ee 
The model describes the motion of a particle in a potential valley
extended along the $x$-axis with quadratic confining potential in the
$y$-direction. The valley is intersected at an angle by the potential
barrier which introduces non-linear coupling between the degrees of
freedom. The process we are interested in is a penetration through the
barrier  of the particle which  comes from the left in a fixed initial
quantum state $\ket{E,E_y}$.  The latter  is characterized by the
total energy $E$ and the energy of oscillations in the
$y$-direction, $E_y=\hbar\omega (n+1/2)$, where $n$ is the occupation
number of $y$--oscillator. Note that we keep $E_y$ fixed in 
the semiclassical limit $\hbar\to 0$, so that $n$ grows to infinity. 
It is shown in \cite{Bezrukov:2003yf} that, for given $E_y$,
 transmission through the
barrier is a tunneling process for total energies $E < E_b(E_y)$,
while at $E>E_b(E_y)$ the transmission proceeds classically. 
The values of $E_b$ 
are considerably higher than the height of the potential
barrier, $E_b(E_y) >V_{max} = 1$. 
The mechanism of transmission changes from direct
tunneling to sphaleron--driven tunneling when the total energy exceeds
a certain critical value $E_c(E_y)$, where $V_{max} < E_c (E_y) <
E_b(E_y)$.  

We start by reviewing the derivation of the formula (\ref{main})
in the ordinary regime of direct tunneling. One considers the
amplitude of transition from the state $\ket{E,E_y}$ at the initial
time moment $t = t_i$ to the state $\ket{{\bf x}_f}$ with definite 
coordinates beyond the barrier at the final moment $t = t_f$. Using
the propagator in the coordinate representation, one writes,
\be
\label{amp}
{\cal A}=\int d{\bf x}_i
\bra{{\bf x}_f}\e^{-iH\Delta t/\hbar}\ket{{\bf x}_i}
\bra{{\bf x}_i}E,E_y\rangle\;, 
\ee
where $\Delta t\equiv t_f-t_i$.
The propagator is given by the semiclassical Van Vleck formula,
\be
\label{VV}
\bra{{\bf x}_f}\e^{-iH\Delta t/\hbar}\ket{{\bf x}_i}
=\frac{\e^{iS^{cl}({\bf x}_i,{\bf x}_f)/\hbar}}{2\pi i\hbar}
\left[\det{\frac{\d^2S^{cl}}{\d{\bf x}_i\d{\bf x}_f}}\right]^{1/2},
\ee
where $S^{cl}({\bf x}_i,{\bf x}_f)$ is the action evaluated on the
classical solution going from ${\bf x}_i$ at $t=t_i$ to ${\bf x}_f$ at
$t=t_f$. It will be important for us that  (\ref{VV}) can be 
derived \cite{Levit:1977}
from the path integral
\be
\label{fi}
\bra{{\bf x}_f}\,\e^{-iH\Delta t/\hbar}\ket{{\bf x}_i}
=\int^{{\bf x}(t_f)={\bf x}_f}_{{\bf x}(t_i)={\bf x}_i}
\!\!\!\!\!
[d{\bf x}(t)]\; \e^{iS[{\bf x}(t)]/\hbar}\;.
\ee  
We take the wave function of the initial state in the form 
$\bra{{\bf x}_i}E,E_y\rangle=\psi(x_i)\Psi(y_i)$, where  
$\psi(x_i)$ is  
the plane wave with the unit flux normalization, and $\Psi(y_i)$ is the
semiclassical expression for the
oscillator eigenfunction.
Evaluating the integral (\ref{amp}) 
in  the saddle--point approximation one obtains,
\be
\label{amp1}
{\cal A}=\sqrt{\frac{\omega}{2\pi D_1}}\e^{i(S^{cl}+B)/\hbar
+i\pi/4}\;,
\ee
where
\begin{align}
&B=\dot{x}_ix_i+\int^{y_i}_{\sqrt{2E_y}/\omega}
dy' \sqrt{2E_y-\omega^2y'^2}\;,
\notag\\
&D_1=\dot{x}_i\dot{y}_i\det{\frac{\d^2(S^{cl}+B)}{\d{\bf x}_i^2}}
\left[\det{\frac{\d^2S^{cl}}{\d{\bf x}_i\d{\bf x}_f}}\right]^{-1}\;.\notag
\end{align}
All the quantities in  (\ref{amp1}) are evaluated on the saddle
trajectory satisfying the initial conditions
\be
\label{bin}
\dot{x}_i^2=2(E-E_y)~,~~~
\dot{y}_i^2+\omega^2y_i^2=2E_y\;,
\ee
where $\dot{\bf x}_i\equiv \dot{\bf x}(t_i)$. 
In
deriving  (\ref{amp1}) we assumed that the saddle configuration is
unique; this is indeed the case for the model
(\ref{Ham}).

The amplitude (\ref{amp1}) is to be inserted into the formula for the
tunneling probability,
\be
\label{tun1}
{\cal P}=\lim_{\Delta t\to\infty}\frac{1}{\Delta t}
\int_{x_f>0}d{\bf x}_f |{\cal A}|^2=
\int dy_f\,\dot x_f\,|{\cal A}|^2\;,
\ee
where  in the second equality we cancelled the divergence of the
integral originating from the infinite region ${x_f\to+\infty}$ by
writing $\frac{1}{\Delta t}\int dx_f=\dot x_f$. 
Note that this operation is legitimate only if the boundary condition 
\be
\label{bf1}
x_f \to + \infty \qquad \mbox{as}\qquad  t_f \to +\infty\;.
\ee
is satisfied.
One substitutes  (\ref{amp1}) into  (\ref{tun1}) and
evaluates the saddle-point integral over $y_f$. One arrives at the
expression (\ref{main}) with 
\be
\label{inst}
F=2\Im(S^{cl}+B)~,~~~
A=\hbar^{1/2}\frac{\omega \dot x_f}{\sqrt{2\pi |D_1|^2 D_2}}\;,
\ee
where $D_2=2\Im\frac{\d^2S^{cl}}{\d y_f^2}$. The saddle--point
equation determines the remaining boundary condition for the tunneling
trajectory,
\be
\label{bf2}
\Im \dot y_f=\Im y_f=0\;.
\ee
Note that reality of the total energy together with
(\ref{bf1}), (\ref{bf2}) imply that $\dot x_f$ is also real. 

One solves numerically the classical equations of motion with the
boundary conditions (\ref{bin}), (\ref{bf1}), (\ref{bf2}) and
discovers \cite{Bezrukov:2003yf} that tunneling
solutions~\footnote{These solutions are defined along a certain 
  contour in the complex time plane which does not, in general,
  coincide with the real time axis.} 
satisfying all the conditions exist only at $E<E_c(E_y)$. 
Still, one can find solutions at
$E_c<E<E_b$ by imposing, instead of  (\ref{bf1}), the reality
condition $\Im \dot x_f=\Im x_f=0$. But the resulting trajectories
never come out from the interaction region: they end up oscillating on
top of the potential barrier. These solutions at $t\to +\infty$ become
precisely the sphalerons we referred to above. 
The formula
(\ref{inst}) for the prefactor is not applicable in this case:
the quantities $D_1$, $D_2$ entering into it describe linear response
of the tunneling solutions to small perturbations of the boundary
conditions and are ill-defined due to
the instability of the solutions.

To deal with this situation, we make the following steps.
First, we introduce a
functional $T_{int}[{\bf x}(t)]$ defined on classical
paths. The choice for $T_{int}[{\bf x}(t)]$ is restricted by three
requirements: 
(a) it must be real and positive--definite on real paths,
(b) it must be finite on paths ending up in the out--region,
(c) it must diverge on paths which stay forever in the interaction
region. Overall, the functional $T_{int}[{\bf x}(t)]$ should roughly
measure the time  spent by the particle in the interaction region. 
For the model (\ref{Ham}) the simplest choice is
$
T_{int}[{\bf x}(t)]=\int dt\, f({\bf x}(t))
$,
where the function $f({\bf x})>0$ vanishes at $x\to \pm\infty$. 

Second, we restrict the path integral (\ref{fi}) to paths
    staying fixed time in the interaction region, $T_{int}[{\bf 
    x}(t)]=\tau$. This eliminates unstable trajectories from the
domain of integration. The full propagator is then recovered by
integrating over $\tau$. This program is realized by inserting the unity
\[
1\!=\!\int\! d\tau\, \delta(T_{int}[{\bf x}(t)]-\tau)\!=\!
\int\! d\tau\!\int_{+i\infty}^{-i\infty}\!\!\frac{id\epsilon}{2\pi\hbar}
\,\e^{\epsilon(\tau - T_{int}[{\bf x}])/\hbar}
\]
into  (\ref{fi}) and changing the order of integration. We obtain,
\be
\label{fie}
\begin{split}
\bra{{\bf x}_f}&\e^{-iH\Delta t/\hbar}\ket{{\bf x}_i}\\
=&\int  d\tau\int_{+i\infty}^{-i\infty}\frac{id\epsilon}{2\pi\hbar}
\e^{\epsilon\tau/\hbar}
\int^{{\bf x}(t_f)={\bf x}_f}_{{\bf x}(t_i)={\bf x}_i}
\!\!\!\!\!\!\!
[d{\bf x}]\,
\e^{i(S[{\bf x}]+i\epsilon T_{int}[{\bf x}])/\hbar}\;.
\end{split}
\ee
One observes that the integral over 
$[d{\bf x}]$ in 
 (\ref{fie}) is the same as in  (\ref{fi}) up to the 
substitution 
\be
\label{Se}
S\mapsto S_\epsilon=S+i\epsilon T_{int}\;.
\ee 
Therefore, one can follow the steps leading  from  (\ref{fi}) to 
 (\ref{amp1}) with the result
\[
{\cal A}=\int d\tau 
\int_{+i\infty}^{-i\infty}\frac{id\epsilon}{2\pi\hbar}
\e^{\epsilon\tau/\hbar}
\sqrt{\frac{\omega}{2\pi D_{1,\epsilon}}}
\e^{i(S^{cl}_{\epsilon}+B_\epsilon)/\hbar+i\pi/4}\;.
\]
Importantly, the semiclassical trajectory ${\bf x}_\epsilon(t)$ here
is a solution to the equations of motion obtained
from the modified action $S_\epsilon$. By construction, it spends 
a finite time interval in the interaction region, and thus is stable. 
The integral over
$\epsilon$ is saturated by the saddle--point $\epsilon(\tau)$ which is
implicitly defined by the condition 
$
T_{int}[{\bf x}_\epsilon]=\tau
$.
The latter follows from  the relation
$
d(S^{cl}_\epsilon+B_\epsilon)/d\epsilon=iT_{int}[{\bf x}_\epsilon]
$.
Note that the saddle-point value $\epsilon(\tau)$ is not necessarily purely 
imaginary; one should be careful to pick the one 
with $\Re\epsilon(\tau)\geq 0$ in order to ensure the 
convergence of the
path integral in  (\ref{fie}). 

One proceeds by substituting the expressions for the amplitude and its
complex conjugate into the tunneling probability and performing
integration over the final coordinates. This leaves the integral over
two interaction times, $\tau$ and $\tau'$, coming from the amplitude
and the complex conjugate amplitude.   
It is convenient to change the integration variables to
$\tau_+=(\tau+\tau')/2$, ${\tau_-=\tau-\tau'}$. The integral over
$\tau_-$ is, again, saturated by the saddle
point; one uses the formula 
$
d(S^{cl}_\epsilon+B_\epsilon-i\epsilon\tau)/d\tau=-i\epsilon
$
in deriving
the saddle-point condition. 
One obtains
\be
\label{tun3}
{\cal P}=\int
d\tau_+
\frac{\omega \dot x_{\epsilon,f}
\sqrt{-d\epsilon/d\tau_+}}{\pi\sqrt{2|D_{1,\epsilon}|^2
D_{2,\epsilon}}}
\e^{-2(\Im(S^{cl}_\epsilon+B_\epsilon)-\epsilon\tau_+)/\hbar},
\ee
where the integral is performed over the real axis, and 
the function $\epsilon(\tau_+)$ is defined by the folowing 
implicit
relation 
\be
\label{tau+}
T_{int}[{\bf x}_\epsilon]+T_{int}[{\bf x}_{-\epsilon}]=2\tau_+\;.
\ee
Note that 
the solution to  (\ref{tau+}) is
generically real for real $\tau_+$:
the equations of motion following from $S_\epsilon$ lead to
${\bf x}_{-\epsilon^*}={\bf x}^*_\epsilon$, which for
real $\epsilon$ implies that the l.h.s of  (\ref{tau+}) is, 
indeed,
real.

So far, we did not refer to the particularities of the
sphaleron--driven tunneling. The step where they become important
is integration over $\tau_+$. For both direct and sphaleron-driven 
tunneling, the integral (\ref{tun3}) is dominated
by the point $\epsilon(\tau_+)=0$ which corresponds to the original
unregularized trajectory. 
In the standard case 
of direct tunneling 
the unregularized trajectory
spends a finite time interval $\tau_+$ in the interaction
region; one obtains
the  expressions 
(\ref{inst}) for the suppression exponent and prefactor
by the saddle--point integration over $\tau_+$. 
In the
sphaleron--driven case the interaction time corresponding to 
$\epsilon=0$
is infinite. 
Thus, the integral is saturated by the end--point of the integration 
interval, $\tau_+\to +\infty$. The tunneling probability in this case is 
determined by the behavior of the integrand in (\ref{tun3}) at 
$\tau_+\to +\infty$, that is, $\epsilon\to+0$. Performing the integration,
one obtains the formula (\ref{main}) with 
\be
\label{sph}
F=\!\lim_{\epsilon\to +0}F_\epsilon\;,~~~
A=\!\hbar^{1/2}\lim_{\epsilon\to +0}
\frac{A_\epsilon}{\epsilon
\sqrt{-4\pi\frac{d\Re T_{int}[{\bf x}_\epsilon]}{d\epsilon}}}\;,
\ee
where $F_\epsilon$ and $A_\epsilon$ are given by  (\ref{inst})
evaluated on the $\epsilon$-regularized solution ${\bf x}_\epsilon$.  
[A subtle point is that 
$F_\epsilon=2(\Im(S^{cl}_\epsilon+B_\epsilon)-\epsilon\tau_+)
=2\Im(S[{\bf x}_\epsilon]+B_\epsilon)$ is to be computed using the
{\em original} action evaluated on the {\em regularized} solution.]
Note an additional factor $\hbar^{1/2}$ in the prefactor compared to
the case of direct tunneling.

Let us summarize our results.
We have derived the following 
prescription for the calculation of the tunneling 
probability in the sphaleron--driven case. First, one
replaces the action of the system with the modified 
action $S_\epsilon$, (\ref{Se}), where $\epsilon>0$. Second, one
finds the tunneling solution ${\bf x}_\epsilon$ of the modified
equations of motion. This solution interpolates between the in-- 
and out--regions 
and is stable. One evaluates its suppression exponent
$F_\epsilon$ and prefactor $A_\epsilon$ using the ordinary ``direct
tunneling'' formulae. Third, one determines the suppression exponent
and prefactor by taking the limit $\epsilon\to +0$ according to
(\ref{sph}). Note that our prescription does not make use of any
particular properties of the illustrative model (\ref{Ham}); it is 
applicable in a large class of models exhibiting the phenomenon of
sphaleron--driven tunneling.

\begin{figure}[b]
\includegraphics[width=0.8\columnwidth]{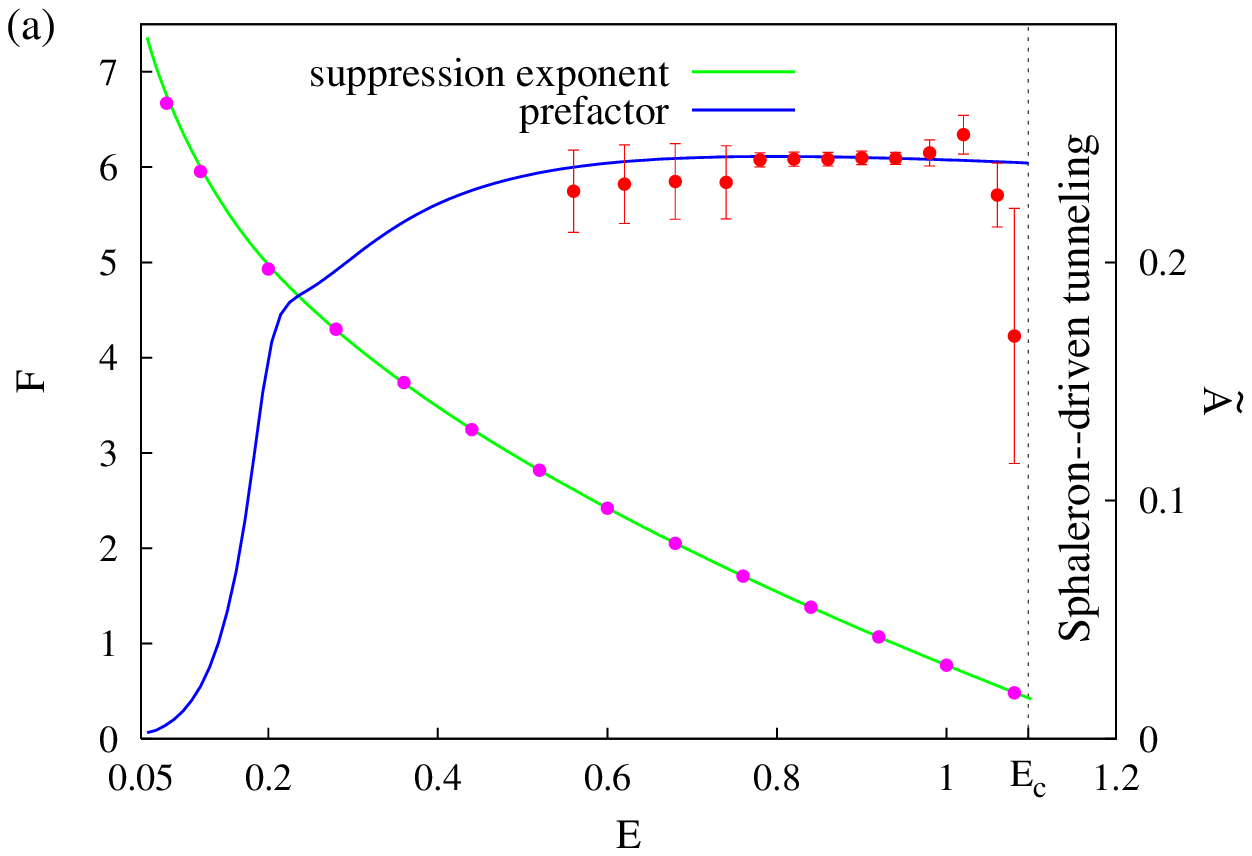}\\
~\\
\includegraphics[width=0.8\columnwidth]{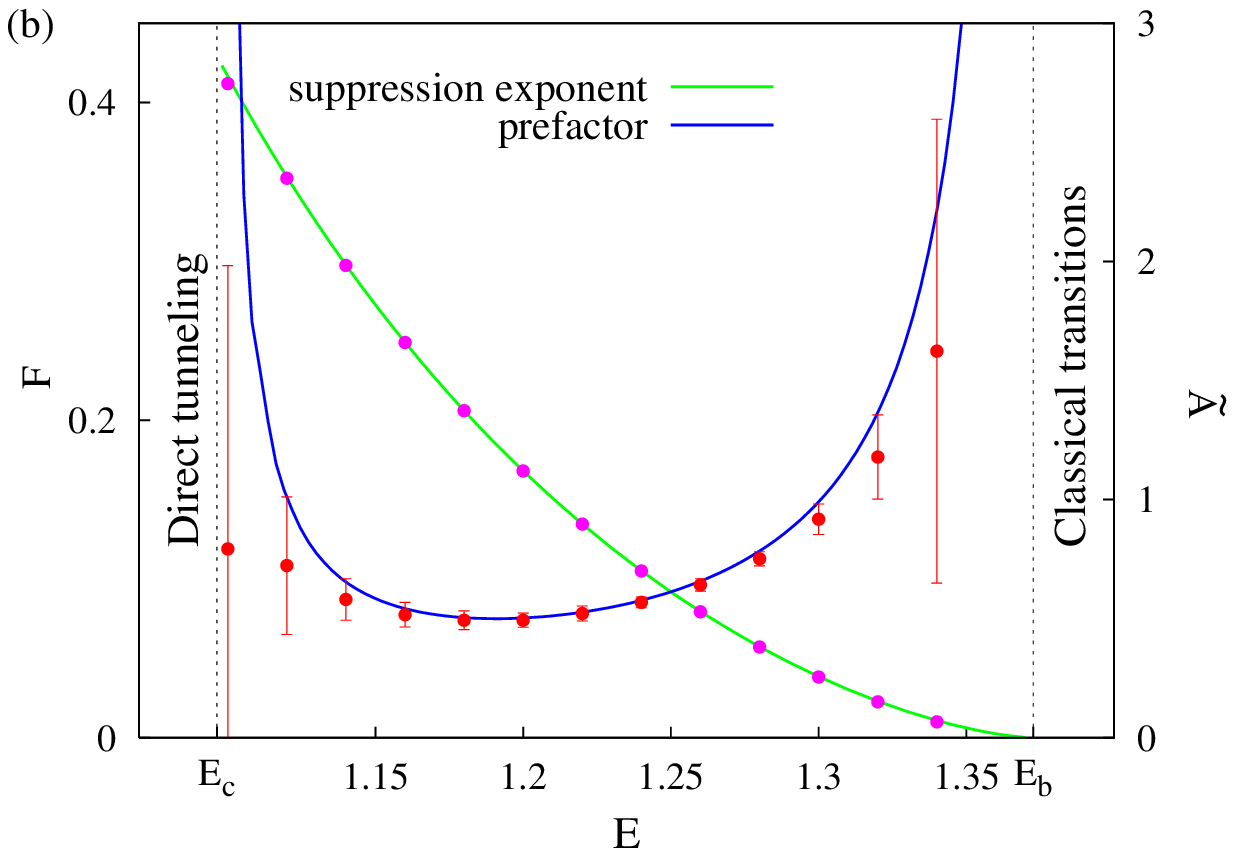}
\caption{Comparison between the semiclassical (lines) and exact
  (points) results for the suppression exponent and prefactor
in the cases of (a) direct and (b) sphaleron--driven tunneling. 
The comparison is performed in the model (\ref{Ham}) with
  $\omega=0.5$, $E_y=0.05$. The error bars represent  the uncertainty
  of the fit (\ref{fit}).}
\label{fig}
\end{figure}
We checked the method of $\epsilon$--regularization by applying it to
the model (\ref{Ham}) and comparing the semiclassical results with the
exact suppression exponent and prefactor extracted from the
numerical solution to  the Schr\"{o}dinger equation. The latter
quantities were
obtained by fitting the dependence of the exact tunneling probability
on $\hbar$ with the formula
\be
\label{fit}
\hbar\log{\cal P}(\hbar)\approx -F+\frac{\gamma}{2}\hbar\log\hbar
+\hbar\log \tilde{A}\;,
\ee
where $\tilde{A}$ is independent of $\hbar$.
For completeness, we also performed the comparison in the regime of
direct tunneling. We set $\gamma=1$ for the direct tunneling and
$\gamma=2$ in the sphaleron--driven case. Figure~\ref{fig} shows the
dependences
\footnote{Due to computational limitations 
we were unable to extract the prefactor $\tilde A$ from the solution
of the Schr\"odinger equation at $E<0.56$.}
 $F(E)$ and $\tilde{A}(E)$ at fixed $E_y=0.05$. 
The semiclassical and
exact quantum--mechanical calculations are in good
agreement.

Note that the quality of the fit (\ref{fit}) becomes worse 
as one approaches
the transition point
$E_c$ between the two tunneling regimes. 
This is a manifestation of
the breakdown of the semiclassical approximation in the vicinity
of this point. 
It is 
caused by the change in the dependence of the prefactor on
$\hbar$.
Indeed,  
$\tilde{A} (E)$ diverges as $E\to E_c+0$, see Fig.~\ref{fig}b, which
contradicts the continuity of the exact tunneling probability. 
However, the size of the
vicinity where the semiclassical approximation breaks down 
vanishes in the limit $\hbar\to 0$.

{\bf Acknowledgements.} We thank to F.L.~Bezrukov, D.S.~Gorbunov 
and
V.A.~Rubakov for discussions. This work was supported in part 
by the RFBR grant 05-02-17363, the Grants of the Russian
Science Support Foundation (D.L. and S.S.), the 
Fellowship of the ``Dynasty'' Foundation (A.P.), and
the EU 6th Framework Marie Curie
Research and Training network "UniverseNet" (MRTN-CT-2006-035863).
The numerical
calculations were performed on the Computational cluster of the
Theoretical division of INR RAS.


\begin{thebibliography}{33}

\bibitem{bohigas}
  O.~Bohigas, S.~Tomsovic and D.~Ullmo, Phys.\ Rep.\ {\bf 223}, 43
  (1993);
  S.~Takada,  P.N.~Walker and M.~Wilkinson, Phys. Rev. A {\bf 52}, 3546
  (1995);
  A.~Shudo, K.S. Ikeda, Phys.\ Rev.\ Lett.\ {\bf 74}, 682 (1995);
  T.~Van~Voorhis, E.J.~Heller, Phys.\ Rev.\ A {\bf 66}, 050501(R) 
  (2002);
 A.D.~Ribeiro, M.A.M.~de~Aguiar, M.~Baranger, Phys.\ Rev.\ E
  {\bf 69}, 066204 (2004);
 S.C.~Creagh, Nonlinearity {\bf 18}, 2089
  (2005); Phys.\ Rev.\ Lett. {\bf 98}, 153901 (2007).
\bibitem{Dembowski}
  C.~Dembowski {\it et al}, Phys.\ Rev.\ Lett. {\bf 84}, 867 (2000).
\bibitem{Hensinger}
  W.K.~Hensinger {\it et al}, Nature (London) {\bf 412}, 52 (2001).
\bibitem{Steck}
  D.A.~Steck, W.H.~Oskay and M.G.~Raizen, Science {\bf 293}, 274
  (2001). 
\bibitem{Miller}
  W.H.~Miller,
  Adv.\ Chem.\ Phys. {\bf 25}, 69 (1974).
\bibitem{recentreviews}
  {\it Tunneling in complex systems}, edited by 
  S.~Tomsovic (World Scientific, Singapore, 1998).
\bibitem{Bezrukov:2003yf}
  F.~Bezrukov and D.~Levkov, arXiv:quant-ph/0301022;
  J.\ Exp.\ Theor.\ Phys.\  {\bf 98}, 820 (2004)
  [Zh.\ Eksp.\ Teor.\ Fiz.\  {\bf 125}, 938 (2004)].
\bibitem{Takahashi:Ikeda}
K.~Takahashi and K.S.~Ikeda, J.\ Phys.\ A {\bf 36}, 7953 (2003);
Europhys. \ Lett. \ {\bf 71}, 193 (2005).
\bibitem{Bezrukov:2003er}
  F.~Bezrukov, D.~Levkov, C.~Rebbi, V.A.~Rubakov and P.~Tinyakov,
  Phys.\ Rev.\ D {\bf 68}, 036005 (2003); 
  D.G.~Levkov and S.M.~Sibiryakov,
  Phys.\ Rev.\ D {\bf 71}, 025001 (2005).
\bibitem{Levkov:2007ce}
  D.G.~Levkov, A.G.~Panin and S.M.~Sibiryakov,
  arXiv:nlin.cd/0701063; 
  arXiv:0704.0409 [quant-ph].
\bibitem{Takahashi:Ikeda2}
K.~Takahashi and K.S.~Ikeda, Phys. \ Rev. \ Lett.\ {\bf 97}, 240403 (2006).
\bibitem{Wilkinson}
Unstable semiclassical solutions in a
different context were studied in
  P.B.~Wilkinson {\it et al.}, Nature (London) {\bf 380}, 608 (1996);
  S.C.~Creagh and N.D.~Whelan, Phys.\ Rev.\ Lett. {\bf 77}, 4975
  (1996); {\bf 82}, 5237 (1998).
\bibitem{Levit:1977}
  S.~Levit and U.~Smilansky, Ann.\ Phys.\ {\bf 103}, 198 (1977).
\end{thebibliography}
\end{document}